# Design heuristics: privacy and portability

Regulation as a feature request

Yasodara Cordova - Harvard University/Unico IDtech

yasodara_cordova@hks.harvard.edu

**ABSTRACT**

The lack of user experience standards in regulations for data privacy and data portability in the health sector increases the cost of leaving a network provider while not protecting the patient's privacy, directly impacting people's health. Furthermore, user in-app options for data sharing and portability in the health sector's applications make it difficult to transfer data between providers while facilitating privacy breaches. Moreover, it leaves users unaware of occasional past unauthorized data access episodes. In this article, we propose an extension for the traditional design heuristics to increase privacy and portability controls for applications that deal with users' personal information based on a benchmark in applications from different sectors and a literature review.

**Keywords:** privacy, portability, heuristics, interface design, experience design.

# INTRODUCTION

The lack of user experience standards in regulations for data privacy and data portability in the health sector increases the cost of leaving a network provider instead of protecting the patient's privacy, directly impacting people's health. User in-app options for data sharing and portability in the health sector's applications make it difficult to securely transfer data between providers, so it facilitates privacy breaches. For example, it leaves users unaware of unauthorized data access while making it difficult for the user to understand the actors using their data.

Users' privacy perception might present nuances, depending on the context. For example, recommendation systems are at the core of social media platforms' operations, and so are the personal data that feeds them. Due to past cases of unethical use of personal data, like the notorious Cambridge Analytica case, there is some awareness of the risks presented by these social networks[1]. For different sectors, users' privacy perception might differ depending on variables like trust in the providers, level of data sensitivity, or even users' trust relationship with the companies providing the services. The health sector, however, collects and shares the user's data despite any trace of the "privacy paradox," or the difference between people's intention to disclose data versus the actual behavior towards data sharing (Norberg et al., 2007), because the health data are collected inside a mandatory relationship of the service providers with the patient. Doctors, nurses, health insurance providers, and other health sector chain members need access to the data to adequately address users' health demands.

Like in social networks, the health sector requires high volumes of personal data to perform, but its use isn't optional. Users of health platforms have little or no option when deciding about using the services that collect data or not.

Given that people seeking medical assistance have to opt-in to give their medical information, the health industry is not necessarily engaged in improving the user interface to retain the user, facilitate their understanding of data sharing options, or aid in data portability issues. **The situation makes it difficult for the patient to 1) access to and control their data and 2) transfer data between different service providers.**

Due to that, patients have little knowledge of who accessed their data, when, or the reason that motivated the access. Simultaneously, the users find it challenging to complete simple tasks, like downloading and securely sharing their data among the chosen network of services or providers, losing access to data when changing services, or having to perform unnecessary clinical exams more than once since they have changed hospitals, doctors or clinics. The effects of this difficulty might impact the user's decision when switching providers or even when deciding whether to go on a medical visit or not. Finally, users' have to repeat their personal information as soon as they log in to new services, creating friction to maintain their data updated and connected with several previously used systems.

---

[1] Díaz Ferreyra, Nicolás, Rene Meis, and Maritta Heisel, 'At Your Own Risk: Shaping Privacy Heuristics for Online Self-Disclosure', 2018 <https://doi.org/10.1109/PST.2018.8514186>

**OBJECTIVES**

The present study aims to improve users' privacy when they use health tech applications and platforms by transferring analyzed features from other sectors to the health sector, using heurístics as standards for actionable recommendations for the industry. This research can also help UX/UI designers guarantee minimal data portability with data privacy for health tech users.

**BACKGROUND**

In 1996, the US introduced federal law requiring the creation of national US standards to protect people's medical information, the Health Insurance Portability and Accountability Act. The act defines medical information as "protected health information" and includes healthcare providers (doctors, clinics, hospitals, nursing homes, and pharmacies), health plans, healthcare clearinghouses, and business associates. A subset of the PHI (protected health information) deals with identifiable health information, protected by the Security Rule, aiming at ensuring "the confidentiality, integrity, and availability of all electronically protected health information[2]." The rule tries to prevent threats to the information's security and prohibited uses or disclosures, creating a compliance chain among the workers involved in processing or accessing that information. In certain situations, any covered entity can use and disclose PHI without an individual's authorization. Among these situations, treatment, payment, and healthcare operations are listed, and the public interest and benefit activities are included in the 12 broad national priority purposes. The HIPAA safeguards include physical, administrative, and technical sectors, the latter being the focus of this article. The technical safeguards[3] cover access control, audit controls, data integrity, authentication, security in data transmission, and mandating encryption/decryption for the latter, in case of ePHI. The regulation focuses on health care access, preventing fraud, providing data for tax-related health provisions, and correctly applying group health insurance requirements.

---

[2] HIPAA Privacy Rule. Accessed on 14 December 2020. Available at: https://www.hhs.gov/sites/default/files/ocr/privacy/hipaa/administrative/combined/hipaa-simplification-201303.pdf

[3] HEALTH INSURANCE PORTABILITY AND ACCOUNTABILITY ACT OF 1996. Public Law 104-191. 104th Congress Available at: https://www.govinfo.gov/content/pkg/PLAW-104publ191/html/PLAW-104publ191.htm

Although the regulation correctly mentions privacy and portability in its technical safeguards, the transformation of these into palpable benefits for most users is still poorly allocated. Another effect of the lack of application of the safeguards in the legislation is the decrease in the consumers' availability of choices, who, without being able to transport their data from one provider to another, might not leave their accredited network. Studying portability cases in companies that provide services based on data retention under the portability rules of the GDPR, such as Spotify, Google, and Facebook, Florez Ramos, and Blind argued that these rules would impact the ability of competitive markets to innovate. This will trigger new technologies to include consumer data acquired from its competitors, which they have called the "exploration-innovation" model. For companies that are not in a competitive environment, the adaptation model would be defined as "exploitation-innovation," not allowing for increasing innovation even with the rules of portability and privacy imposed by the GDPR. The US healthcare system is inefficient and less competitive than it should be[4] to incentivize innovation, so there is evidence that companies are not getting the right incentives to improve customer portability and privacy. Practices among users, however, require adaptation of the system when patients use applications outside the network, such as e-mail and text messages, to circumvent the difficulties presented by the closed and inefficient systems of the health tech providers. The consequences of using such systems are dreadful for the users, for it threatens their privacy and might result in data breaches.

The practice that can improve the aspects mentioned earlier is the creation of more efficient interfaces to allow users to exchange information between participants in the ecosystem in a secure way, or download their data to share securely with other actors. The portability rules will affect the design of the applications (Kolb, 2019), if not by the competition incentives, by the user's adoption of out-of-the-network technologies for porting data (like WhatsApp messages, for example). Recently, the importance of text messaging increased due to COVID-19 lockdown measures. HIPAA regulations for these matters sparked the development of new applications to facilitate communications between providers and patients, as seen in applications like TigerText, Zinc, QliqSoft, and Spok Mobile. These applications follow HIPAA regulations for text messaging when dealing with PHI: unique user IDs, authentication via password or pin, smart card or token, and even authentication via facial recognition. Further, HIPAA regulations ensure that entities have processes to create access rights in emergency cases, automatic logoff of users, and the protection of data integrity when using text to send and receive information. Even with the rules for ensuring users' privacy, the HIPAA regulations for text messages do not imply data portability features. For that, HIPAA regulations should incentivize the creation of interfaces that allow data portability while keeping PHI safe, guaranteeing users' privacy.

---

[4] Avraham, R., 2011. Clinical practice guidelines: the warped incentives in the US healthcare system. American journal of law & medicine, 37(1), pp.7-40.

**DESIGN HEURISTICS, PRIVACY, AND PORTABILITY**

Heuristics are procedures or rules proposed to acquire and discover solutions to a specific problem. Heuristics are developed by implementing guidelines for knowledge acquired in previous analyzes of practices in a given sector. The method of proposing heuristics to solve problems in products is not new. Still, it is particularly useful in Human-Computer-Interaction, or HCI, to define new practices based on experimentation.

For example, Jakob Nielsen's ten usability heuristics for user interface design defines guidelines for implementing digital interaction between humans and machines on screens with minimal requirements for comfort and accessibility. These guidelines have been used widely by the technology industry. Nielsen's heuristics are basic, such as "visibility of system status", or "Consistency and standards" and "help users recognize, diagnose, and recover from errors,"; but at the same time, the rules are crucial for maintaining features that empower the users and set standards for the industry.

The "Privacy by Design" 7 Foundational Principles (Cavoukian) can be considered an inspiration for building heuristics that help create user interfaces that preserve privacy rights. The seven principles are 1) privacy as the default; 2) proactive, not reactive; preventative, not remedial; 3) privacy embedded into the design; 4) full functionality – positive-sum, not zero-sum; 5) end-to-end security – lifecycle protection; 6) visibility and transparency; and 7) respect for user privacy. The principles of Privacy by design are general, guiding the implementation to preserve users' privacy in a broad context. The principle of data portability is not seen in the Privacy by Design principles by default. Still, product designers can infer new vocabularies to deal with data portability from each principle's definitions. For example, the author describes "respect for user privacy" principles as the fact that "individuals shall be provided access to their personal information and informed of its uses and disclosures."

Data portability per se can be considered a concept that is additional to privacy, nor derived from it. Data portability can exist without privacy, as the other way round.

The ideal scenario is that the safeguards for granting the right to privacy and data portability go hand in hand, implemented through extensive user-centric approaches.

**BENCHMARK**

For this study, we have considered applications that transform users' perceptions of non-tangible resources, comparing personal data to these resources. The visual representation of data can be compared to the visual representation of electricity, virtual currency, and loyalty points because data is not tangible, much as the items described. The efforts to allow users to control these items start with the user experience, in which designers build the interactions with possible controls from the ground up.

Even though adopting heuristics for privacy and portability is not the objective of the teams deploying these solutions, the concepts are useful for sectors with similar characteristics.

1) **Loyalty programs:**

Literature in the field shows that the visual representation of "points" in reward systems in loyalty applications can increase the value perception over the "points" accumulated (*Youjae Yi, Hoseong Jeon, 2003*). Data transfers and collection can become tangible actions for the user if portability is implemented in products as part of the product strategy, considering that data is a non-tangible artifact.

In the same way, data transfers are implicit in loyalty points transfers. Applications to allow the consumer to transfer points between business providers are documented in patents, such as the "US 7.467,096 B2", partially described in the figure next.

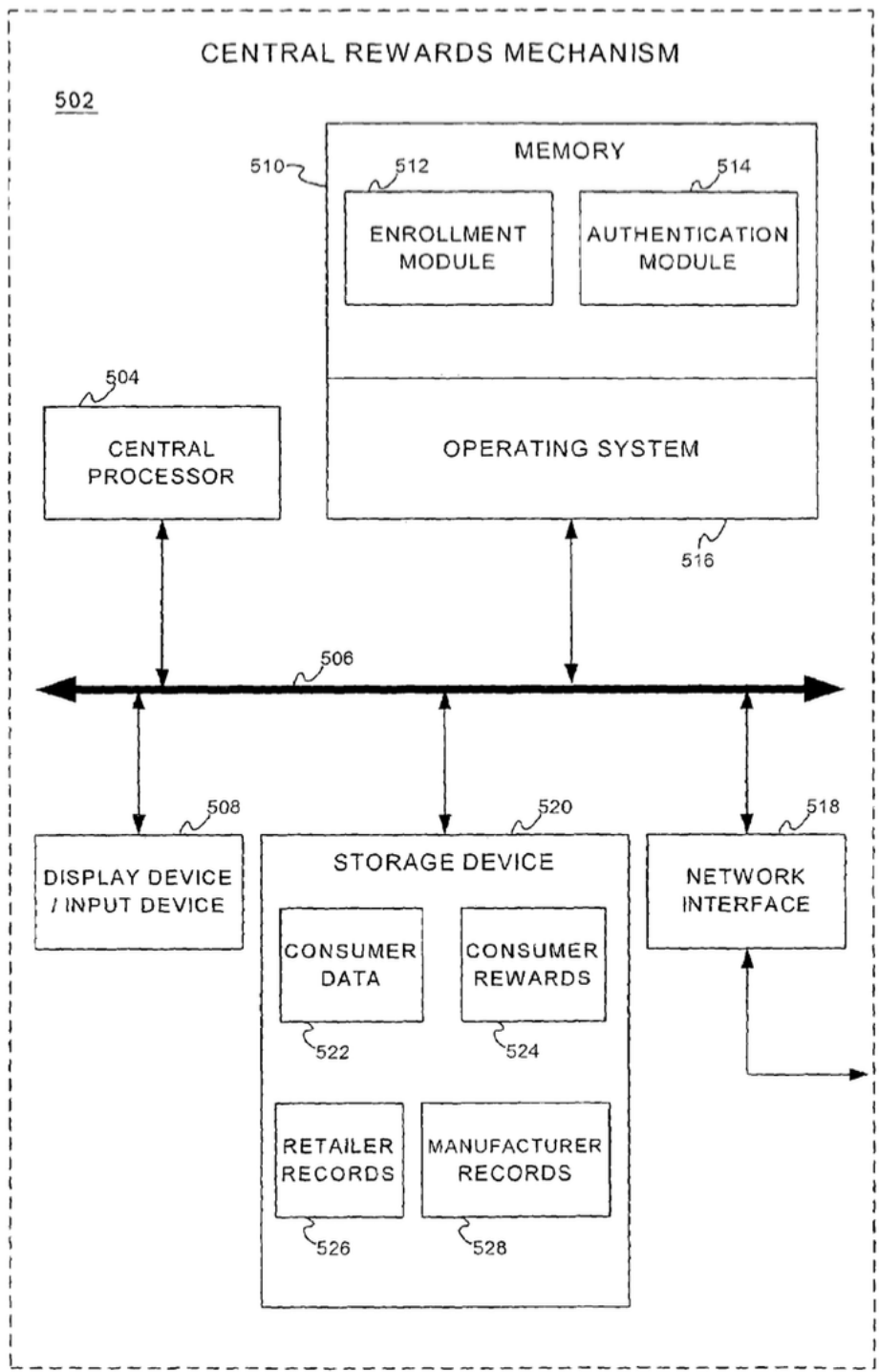

*Figure 1: data transfers in loyalty programs, patent n. US 7.467,096 B2.*

Crucial elements appear when analyzing patents from the sector: receiving a transfer request, accessing and analyzing the total number of loyalty points, determining if any rules exist for restricting or limiting the transfer of points, and adding points to the users' data interface. The actions impacted by these definitions result in user choices regarding their own artifacts. These elements, once invisible to the users, are now available as value-added and can be exchanged and viewed as so.

Visual representations in loyalty apps make it possible for users to exchange their "points" for goods or services.

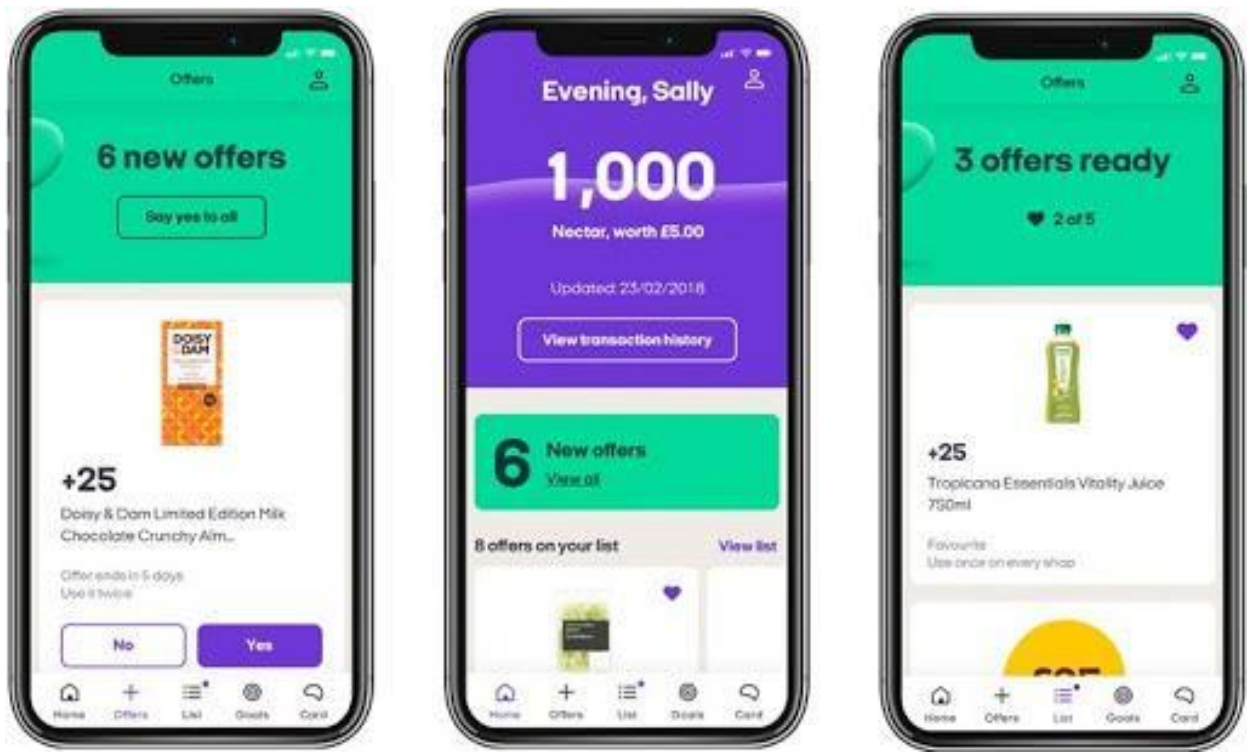

Figure 2 is an example of the users' possibilities in loyalty programs' applications.

Figure 2 brings the visual implementation, or the user interface, for handling data represented by loyalty points. Data requests and offers can be realized without friction, ensuring the data is safe and kept in the user's control.

**2) Electricity control**

Applications designed to allow the user to control the energy generated in their solar panels generally enable users to manage and safely control electricity production in their private grid. Previous analysis of use cases where the apps were introduced in remote communities in developing countries demonstrates that the visual representation of the solar panels' energy helps users control usage and transference of the electricity (Jhunjhunwala et al., 2016). These objectives are achieved by offering a visual representation of the energy generated and possibly returned to the common grid or the amount spent in each homeroom. The features presented in such apps give the user an idea of how much electricity they have and allow them to evaluate consumption or transfer. Such actions would be useful when dealing with PHI, as users need a visual idea about the various aspects of medical information they have given to each platform.

Solar panels' production might not be subject to great scrutiny regarding user privacy, but it represents private resources that the user might not be comfortable making public.

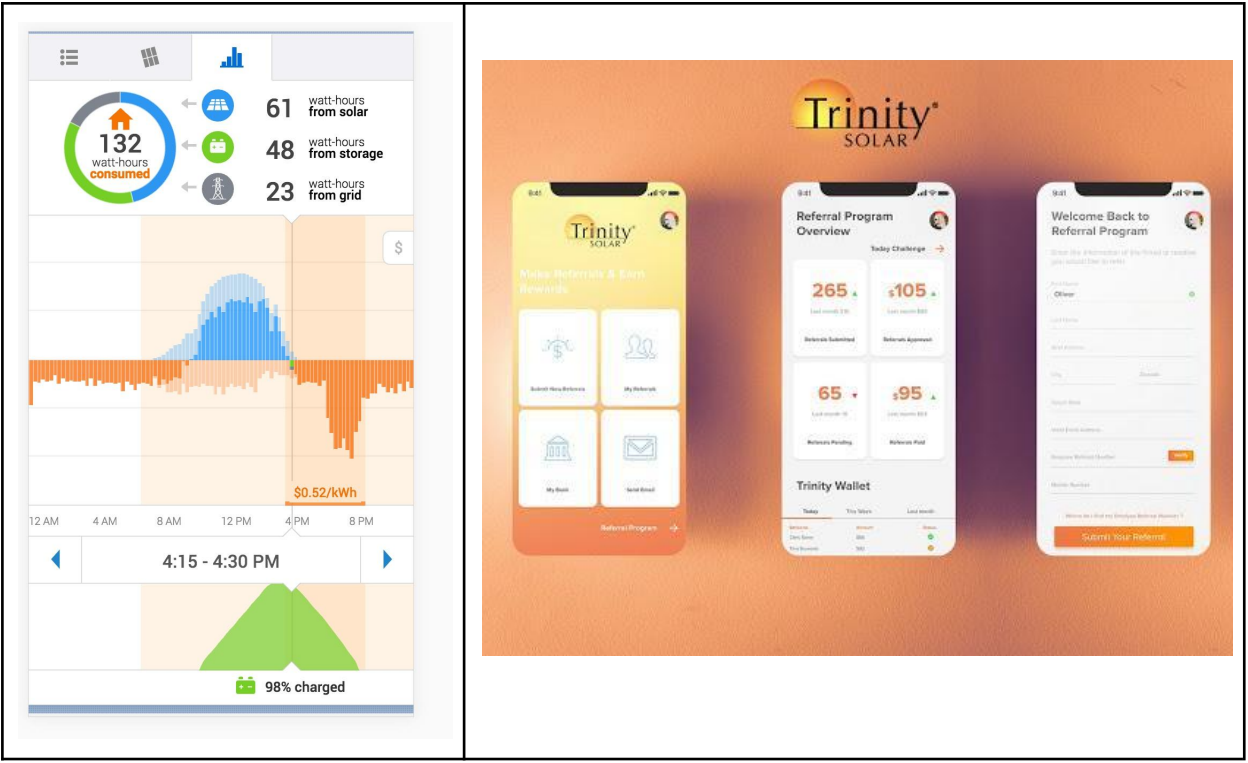

*Figure 3: examples of applications that offer a quantified idea of the energy generated by solar panels*

3) **Financial sector:**

Improvements in how the financial sector handles users' data have received a boost from fintech. Startups are born without the traditional holders of banks and institutions and offer more options so that users can control their financial resources.

With the advent of fintech, control options have been added, but also recommendation options, which act as advisors to the user.

In the financial sector, the applications must offer interfaces where the user can exchange data between institutions and ensure that their information is safe. Users want to have the possibility of taking their funds from one institution to another and, at the same time, paying via social networks, payment apps, etc.

The sector has been driven by the adoption of the open banking standards, created to allow data portability between institutions, aiming to increase the sector's competitiveness, improving the possibilities for the user.[5]

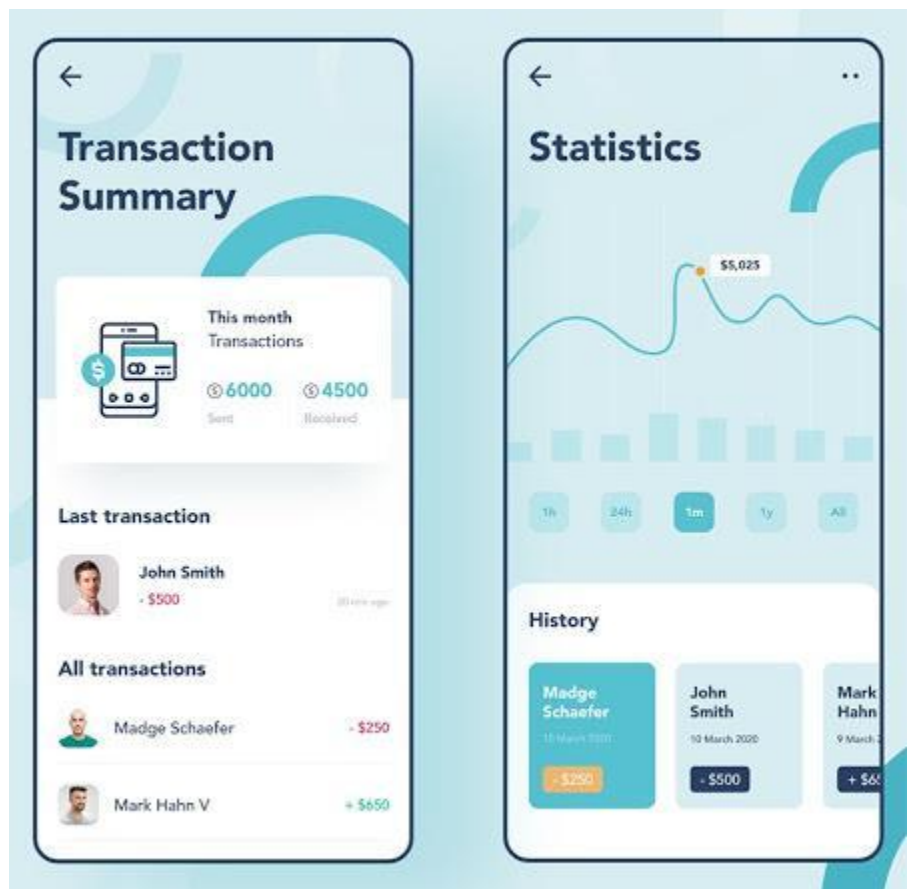

*Figure 4: a generic template for creating an application for the financial sector*

---

[5] Phaneuf, A. (no date) How open banking and bank APIs are boosting fintech growth, Business Insider. Available at: https://www.businessinsider.com/open-banking-api-trends-explained (Accessed: 13 December 2020).

**RESULTS**

Lawrence Lessig coined the expression "code is law" in his book "Code and Other Laws of Cyberspace" when discoursing the regulatory power contained in written software. The Human-Computer Interaction, as the tool that shapes the rules written in code for appropriate contact with humans, can be considered the "law enforcement tool" of cyberspace. In the same way that platforms like Facebook only offer certain liberties to their users, shaping the user interface according to the data collection purposes (previously modeled by data models and software architecture), applications shape the user's possibilities inside cyberspace not just by reflecting the software that is behind the screen, but embedding values and product objectives regarding data collection.

Based in the previous analysis and in the principles already exposed in the article, the proposed privacy heuristics are below.

**Privacy Design Heuristics**

1. **All data should be portable:**

Users should be able to download their data for archival purposes and transfer data from one health insurance platform to another, following global standards;

2. **Users should see "where" their data are:**

Users should be capable of identifying, both geographically and institutionally, where their data is being stored;

3. **Users should know how much and which data is being transferred**

Every user should be able to quantify their personal information, and also identify the size of files, as well as the format.

4. **Users should be able to delete their data**

User's should be capable of deleting their data from health providers' platforms.

5. **Users should be aware of the access permissions regarding their data**

User's should be able to receive notifications when their data is accessed and identify who accessed, why, when, and who authorized it.

6. **Users should be able to revoke, grant, and dispute access to personal data**

The heuristics proposed by this article are not final. Health information is one of the most delicate types of data, and at the same time it is extremely valuable for public policies purposes, administration and innovation in the sector. Hence, data availability and integrity becomes a competitive advantage for the ecosystem. Health information should transit between doctors, nurses, clinics, and all types of participants in the ecosystem but securely and always preserving users' privacy.